\documentclass[aps,pre,twocolumn,showpacs,groupedaddress,nofootinbib]{revtex4-1}  
\usepackage{graphicx}  
\usepackage{dcolumn}   
\usepackage{bm}        
\usepackage{amsmath, amsthm, amssymb}
\usepackage{bbold} 
\usepackage{mathrsfs}
\usepackage{array}
\usepackage[usenames]{color}
\usepackage{soul} 
\usepackage{ulem} 
\begin{document}
\normalem

\title{Percolation on Lieb lattices}
\author{W. S. Oliveira}
\altaffiliation[Present address: ]{Instituto de F\'\i sica, Universidade Federal do Rio de Janeiro Cx.P. 68.528, 21941-972 Rio de Janeiro, Rio de Janeiro, Brazil}
\author{J. Pimentel de Lima}
\affiliation{Departamento de F\'\i sica, Universidade Federal do Piau\'\i, 64049-550 Teresina, Piau\'\i, Brazil}
\author{N. C. Costa}
\affiliation{Instituto de F\'\i sica, Universidade Federal do Rio de Janeiro Cx.P. 68.528, 21941-972 Rio de Janeiro, Rio de Janeiro, Brazil}
\author{R. R. dos Santos}
\affiliation{Instituto de F\'\i sica, Universidade Federal do Rio de Janeiro Cx.P. 68.528, 21941-972 Rio de Janeiro, Rio de Janeiro, Brazil}
\begin{abstract}
We study site- and bond-percolation on a class of lattices referred to as Lieb lattices. 
In two dimensions the Lieb lattice (LL) is also known as the decorated square lattice, or as the CuO$_2$ lattice; in three dimensions it can be generalized to a layered Lieb lattice (LLL) or to a perovskite lattice (PL).
Emergent electronic phenomena, such as topological states and ferrimagnetism, have been predicted to occur in these systems, which may be realized in optical lattices as well as in solid state. 
Since the study of the interplay between quantum fluctuations and disorder in these systems requires the availability of accurate estimates of geometrical critical parameters, such as percolation thresholds and correlation length exponents, here we use Monte Carlo simulations to obtain these data for Lieb lattices when a site (or bond) is present with probability $p$. 
We have found that the thresholds satisfy a mean-field (Bethe lattice) trend, namely that the critical concentration, $p_c$, increases as the average coordination number decreases; our estimates for the correlation length exponent are in line with the expectation that there is no change in the universality class.
\end{abstract}


\maketitle

\section{Introduction}

\begin{figure*}
\centering
\includegraphics[width=\textwidth]{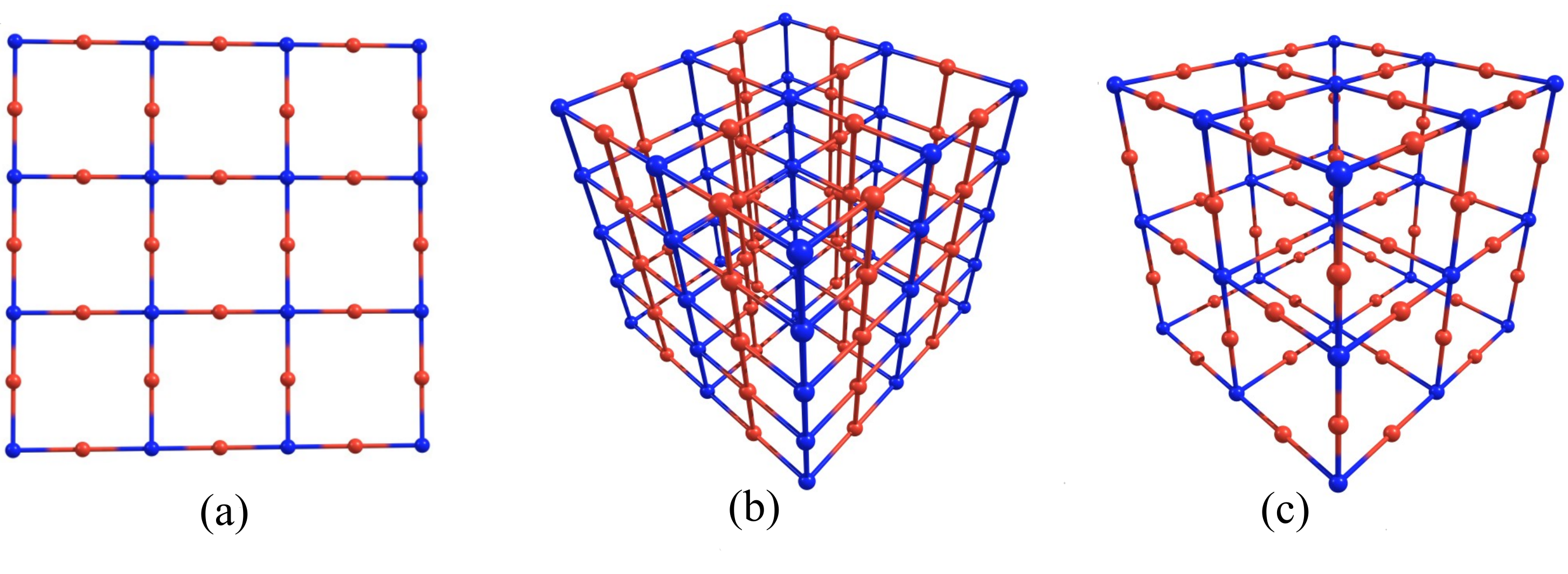}
\caption{The two-dimensional Lieb lattice (a), and its three-dimensional extensions: (b) the layered Lieb lattice, LLL,  and (c)~the perovskite lattice, PL. 
}
\label{fig:lattices}
\end{figure*}

Fascinating electronic properties such as topological states \cite{Weeks10,Beugeling12,Jiang19,Jiang20} and ferrimagnetism \cite{Lieb89,*Lieb89err,Noda15,Costa16,Costa18,Yarmohammadi18} have been highlighted in connection with lattice geometries leading to flat (or dispersionless) bands in the non-interacting limit. 
In two dimensions, for instance, flat bands appear in the so-called Lieb lattices, also known as  CuO$_2$ lattices, where the four-coordinated sites represent Cu atoms and doubly-coordinated sites represent O sites; see Fig.\,\ref{fig:lattices}(a). 
Robust ferrimagnetism in the ground state emerges when fermions are allowed to interact via an on-site repulsive Hubbard coupling \cite{Costa16}. 
The possibility of producing this geometry with ultracold atoms in optical lattices \cite{Shen10,Taie15,Schafer20} 
has stirred even more interest, due to the unprecedented experimental control and tunability of parameters in these systems, such as interaction strength and particle density \cite{Bloch08, Esslinger10}. 

Further, it is well known that in the strong coupling limit and at half-filling (i.e.\,one fermion per orbital, on average) the charge degrees of freedom of the  Hubbard model are frozen, so that the dynamics is dominated by the spin degrees of freedom, whose scale of interaction is set by their exchange coupling, $J\sim W^2/U$, where $W\, (\ll U)$ is the bandwidth. 
For instance, on hypercubic lattices the ground state is a Mott insulating antiferromagnet. 
Since the ground state (or even  low temperature) properties in these systems are dominated by quantum fluctuations, an issue which has been the subject of scrutiny over the years is the interplay between quantum fluctuations and random, geometric disorder.
Indeed, after a long lasting debate it was finally settled that the purely geometric percolation threshold \cite{Stauffer94} coincides with the onset of antiferromagnetic order in the dilute Heisenberg model \cite{Sandvik02}. At any rate, one may take the classical percolation threshold as a lower bound for magnetic order in the Heisenberg model. This is therefore a clear indication that one must have at our disposal accurate estimates for the critical concentrations in these purely geometrical problems.
\color{black}
Away from the strong coupling limit of the Hubbard model it has been established that the itinerant character of the particles strongly influences the threshold value for magnetic (or superconducting) order in the ground state: it can be either larger (as for the square lattice \cite{Ulmke98b,Litak00,Hurt05,Mondaini08,Pradhan18}) or smaller (as for the 2D Lieb lattice \cite{Lima20}) than the geometrical (percolation) threshold. 
Dilution in these cases is implemented by switching off the on-site interaction, $U$, on a fraction $x$ of sites, or, in a notation more akin to the present context of percolation, by rendering a site occupied (or active) on a fraction $p\equiv 1-x$ of sites. 

Interesting extensions of the Lieb lattice to three dimensions can be obtained as follows:
one can either pile up layers of 2D Lieb lattices, as shown in Fig.\,\ref{fig:lattices}(b), or one can form a `perovskite lattice', in which a lattice site is introduced along the $c$-axis halfway between two Lieb layers, so that each face of the cube looks the same, as in Fig.\,\ref{fig:lattices}(c). 
It is important to note that the 3D layered Lieb lattice (LLL) does not display a flat band \cite{Noda15}, while the perovskite lattice (PL) displays a doubly degenerate flat band \cite{Weeks10}. 
Indeed, in two dimensions the flat band emerges due to the existence of localized states associated with the `O sites'. 
However, when we pile these 2D lattices to form the LLL, the electrons become delocalized along the $c$-axis; by contrast, when we form the PL there is no delocalizing channel available, and the flat band is preserved.

In view of the above discussion related to the disordered Hubbard model in 2D lattices, this difference immediately raises the question of how the interplay between itinerancy and lattice geometry is translated to these 3D Lieb lattices. 
In this context, it is therefore crucial to have at hand accurate estimates for the percolation thresholds in these lattices.
With this in mind, here we use Monte Carlo simulations to determine the critical concentrations for site- and bond-percolation on Lieb lattices. Although in principle one should not expect any new universality classes by changing the lattice structure at fixed spatial dimensionality, we also present estimates for the corresponding correlation length exponents in order to confirm this.
The layout of the paper is as follows: In Sec.\,\ref{sec:meth} we briefly describe the way the disorder configurations are generated, and how finite size effects are used to our benefit to perform extrapolations to the infinite lattices.
In Sec.\,\ref{sec:results} we present and discuss the results obtained, and Sec.\,\ref{sec:conc} summarizes our findings.


\color{black}

\section{Methodology}
\label{sec:meth}

\begin{figure}
\includegraphics[scale=0.50]{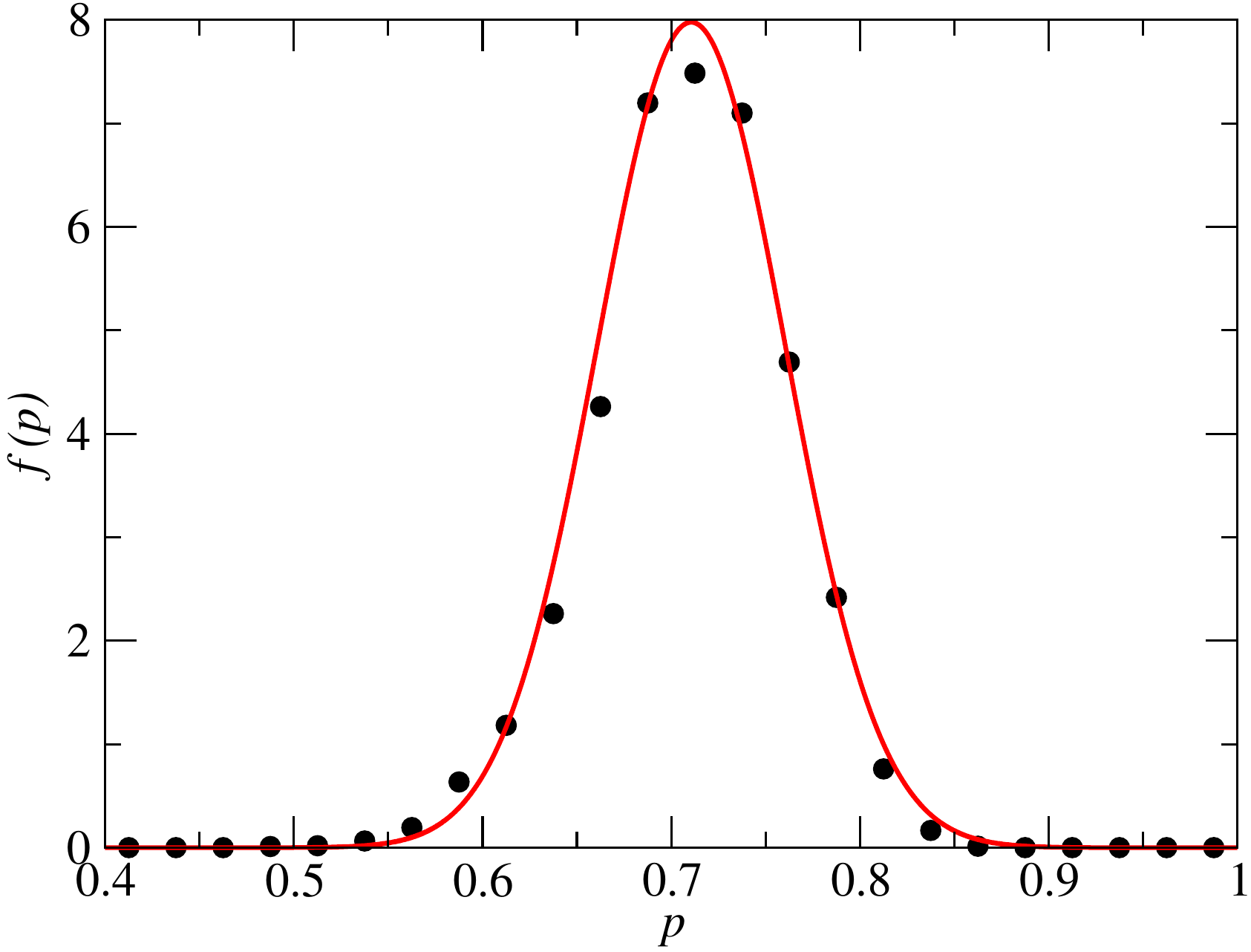}
\centering
\caption{Distribution of concentrations for $M$\,=\,5,000 randomly generated percolating two-dimensional Lieb lattices with $L=20$: the full (red) curve is a gaussian fit through the data points.}
\label{fig:fofp}
\end{figure}

We generate a disorder configuration by randomly occupying the sites (or bonds) of a lattice, and test whether the configuration percolates or not. 
This test is carried out through the Hoshen-Kopelman algorithm \cite{Hoshen76}, according to which  the lattice is scanned while clusters of connected nearest neighbor sites acquire labels; if later in the scanning process a cluster is found to be connected to another cluster, the labels are merged by keeping the smallest one amongst the two clusters.
When the whole scan is completed, and if the cluster labels in opposite extremes of the lattice are the same, the system is said to percolate; we discard non-percolating configurations. 
In this way, we generate an ensemble of $M$ percolating lattices, with a distribution of concentrations peaked near the percolation threshold; a typical example of such distribution, $f(p)$, is shown in Fig.\,\ref{fig:fofp}.
We evaluate the average value of $p$ in this ensemble and its standard deviation, which respectively become our estimate for $p_c(L)$ and its error bar. 

According to finite-size scaling (FSS) theory, these $p_c(L)$ are expected to behave as \cite{Fisher71,Barber83,Stauffer94}
\begin{equation}
	p_{c}(L)=p_{c}+AL^{-1/\nu}, \quad L\gg 1,
\label{eq:FSS}
\end{equation}
where $p_{c}$ is the percolation threshold in the thermodynamic limit,  $A$ is a nonuniversal 
amplitude, and $\nu$ is the correlation length exponent. 
One should note that the lattices considered here have bases, so that the relation between $L$ and the number of sites is not the same as for hypercubic lattices; see Sec.\,\ref{sec:results}.
Fitting our data for $p_c(L)$ to this form provides estimates for $p_{c}$ and $\nu$ in each of the different cases described below.
\color{black}

\section{Results and discussions}
\label{sec:results}

We first gauge the accuracy of our method by examining site- and bond-percolation on the standard square and (simple) cubic lattices.  
For the square lattice, we have used between $M=10,000$ runs for the smallest $L\times L$ lattices, $L=24$, and $M=3,000$ for the largest, $L = 160$, lattices; for the cubic lattice, we have used between $M=8,000$ runs for the smallest $L\times L\times L$ lattices, $L=10$, and $M=800$ for the largest, $L = 48$, lattices.
The extrapolated values for $p_c$ and $\nu$ obtained through the FSS ansatz, Eq.\,\eqref{eq:FSS}, are displayed in Table \ref{tab:pcnu}, from which we see that our results compare very well with previous estimates (simulations and series expansions) or exact results, where available.
This gives us confidence that the procedure outlined above is indeed able to provide accurate estimates for the different Lieb lattices considered here.

\begin{table}\footnotesize
\begin{tabular}{c|c|c|c|c}
Lattice & s or b  & $p_c$                    & $\nu$ & Comments  \\
\hline
\hline
square & s         & $0.5920(5)$ & $1.30(6)$ & $p_c=0.59274598(4)^a$; $\nu=4/3^b$\\
	 & b         & $0.4993(5)$ & $1.29(5)$ & $p_c=1/2^c$; $\nu=4/3^b$\\
\hline
Lieb & s         & $0.7396(5)$ & $1.35(4)$ & \\
	 & b         & $0.6438(3)$ & $1.30(5)$ & \\
\hline\hline
simple  & s         & $0.3118(5)$ & $0.90(7)$ & $p_c=0.311681(13)^{d,e}$\\
cubic&	       &		     &			& $\nu=0.88(2)^{e,f}$\\
	& b         & $0.2484(5)$ & $0.89(6)$ & $p_c=0.24881182(10)^e$ \\
	&	       &		     &			& $\nu=0.8764(15)^e$\\
\hline
LLL & s         & $0.3919(5)$ & $0.86(5)$ & \\
	 & b         & $0.3338(5)$ & $0.88(4)$ & \\
\hline
PL & s         & $0.5225(5)$ & $0.89(3)$ & \\
	 & b         & $0.4010(5)$ & $0.87(4)$ & \\
\hline
\end{tabular}
\caption{Estimates obtained for the critical concentration, $p_c$, and for the correlation length exponent, $\nu$; see text.
LLL and PL respectively stand for layered Lieb lattice and perovskite lattice; see Fig.\,\ref{fig:lattices}. 
The numbers in parentheses are the uncertainties in the last digit(s). 
Notes: $^a$Ref.\,\cite{Lee08}; $^b$Exact, Ref.\,\cite{denNijs79}; $^c$Exact, Ref.\,\cite{Stauffer94}; 
$^d$Ref.\,\cite{Ballesteros99}; $^e$Ref.\,\cite{Wang13}; $^f$Ref.\,\cite{Gaunt83}.} 
\label{tab:pcnu}
\end{table}

\begin{figure}[t]
\includegraphics[scale=0.5]{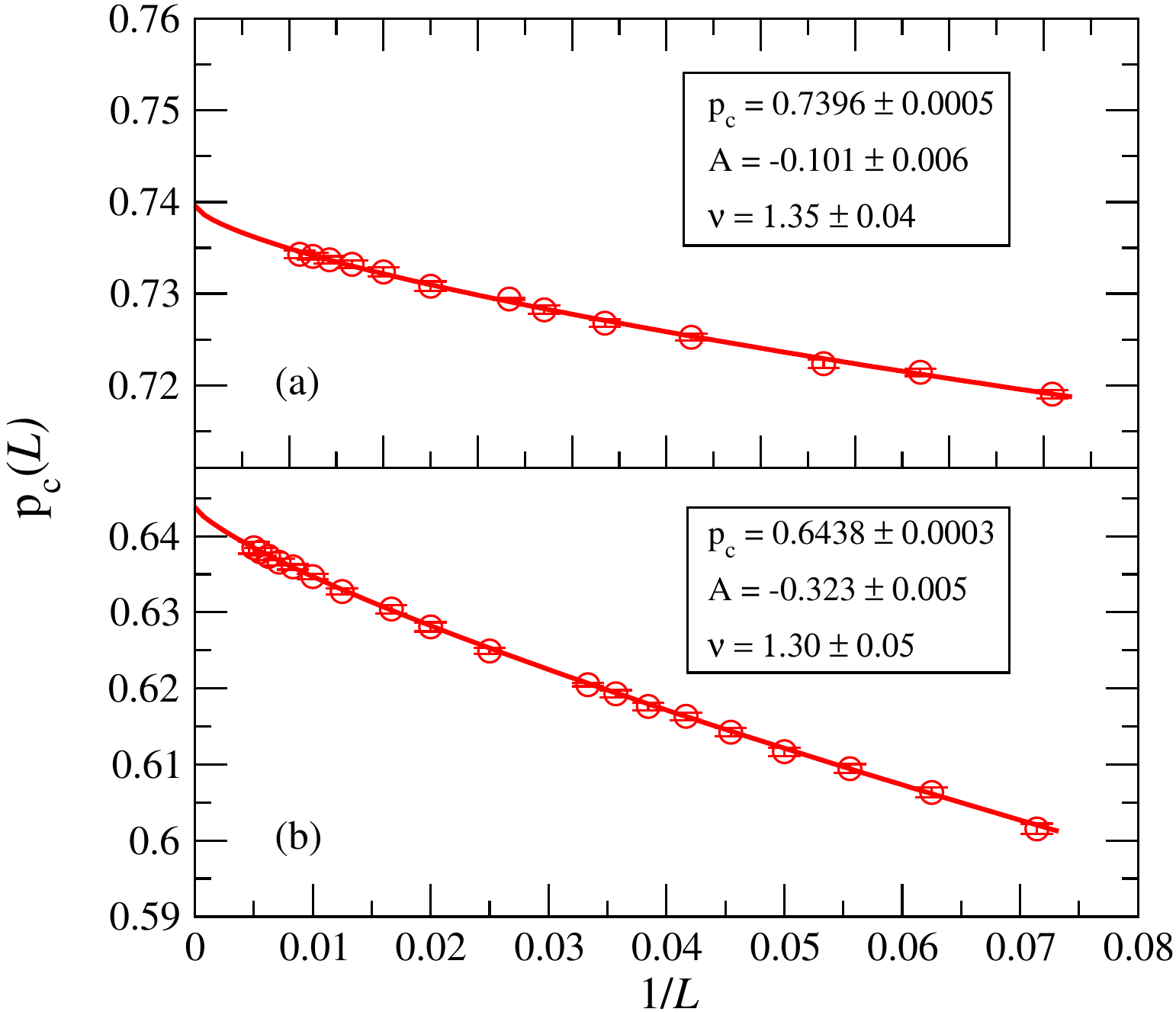}
\centering
\caption{Site (a) and bond (b) percolation thresholds, $p_c(L)$, as functions of the inverse linear lattice size, $1/L$, for the two-dimensional Lieb lattice. 
The (red) curves are fits of the data to the FSS scaling form, Eq.\,\eqref{eq:FSS}, from which one infers the $L\to\infty$ value for $p_c$ and $\nu$, and the amplitude $A$; the error bars supplied in the box arise from the fitting process.
}
\label{fig:lieb2d}
\end{figure}

For the two-dimensional Lieb lattice, we note that the unit cell contains 3 sites (one `Cu-like' and two `O-like'), so that the linear lattice size, $L$, actually contains $2L$ sites; that is, the computational effort is measured by the total number of sites, while the relevant FSS parameter is $L$. 
Accordingly, we have used between $M=10,000$ runs for the smallest $L\times L$ lattices, $L=5$, and $M=1,000$ for the largest, $L = 90$, lattices.
The FSS plots for the average thresholds for site- and bond-percolation are shown in Figs.\,\ref{fig:lieb2d}(a) and (b), respectively.
The extrapolated data are shown in Table \ref{tab:pcnu}, from which we see that the thresholds for the 2D Lieb lattice are higher than the corresponding ones for the square lattice. 
Indeed, in the case of site-percolation the O-sites have smaller connectivity than the Cu-sites, so that one needs a larger overall concentration of sites to make up for this; a similar argument holds far the case of bond-percolation.
Further, this is also consistent with the mean-field result (see, e.g.\,Ref.\,\cite{Stauffer94}), $p_c=(z-1)^{-1}$, if we interpret $z$ as an average coordination number.
At this point we should also note that the site-percolation problem on the 2D Lieb and perovskite lattices is connected to the site-bond percolation problem on the square and simple cubic lattices, respectively. 
In this percolation problem, each site is present with probability $p_s$, and each bond is independently present with probability $p_b$; thus, when $p_s=p_b$ one has the site-percolation problem on the Lieb or perovskite lattices. 
Phase diagrams $p_b\times p_s$ have been obtained for several lattices in Ref.\,\cite{Tarasevich99},  but thresholds for the cases $p_s=p_b$ are not available in accurate tabular form.
Table \ref{tab:pcnu} also shows that the correlation length exponent, $\nu$, is the same, within error bars, as for the square lattice, thus confirming our expectation that the modified geometry of the Lieb lattice is unable to change the universality class. 

\begin{figure}[t]
\includegraphics[scale=0.5]{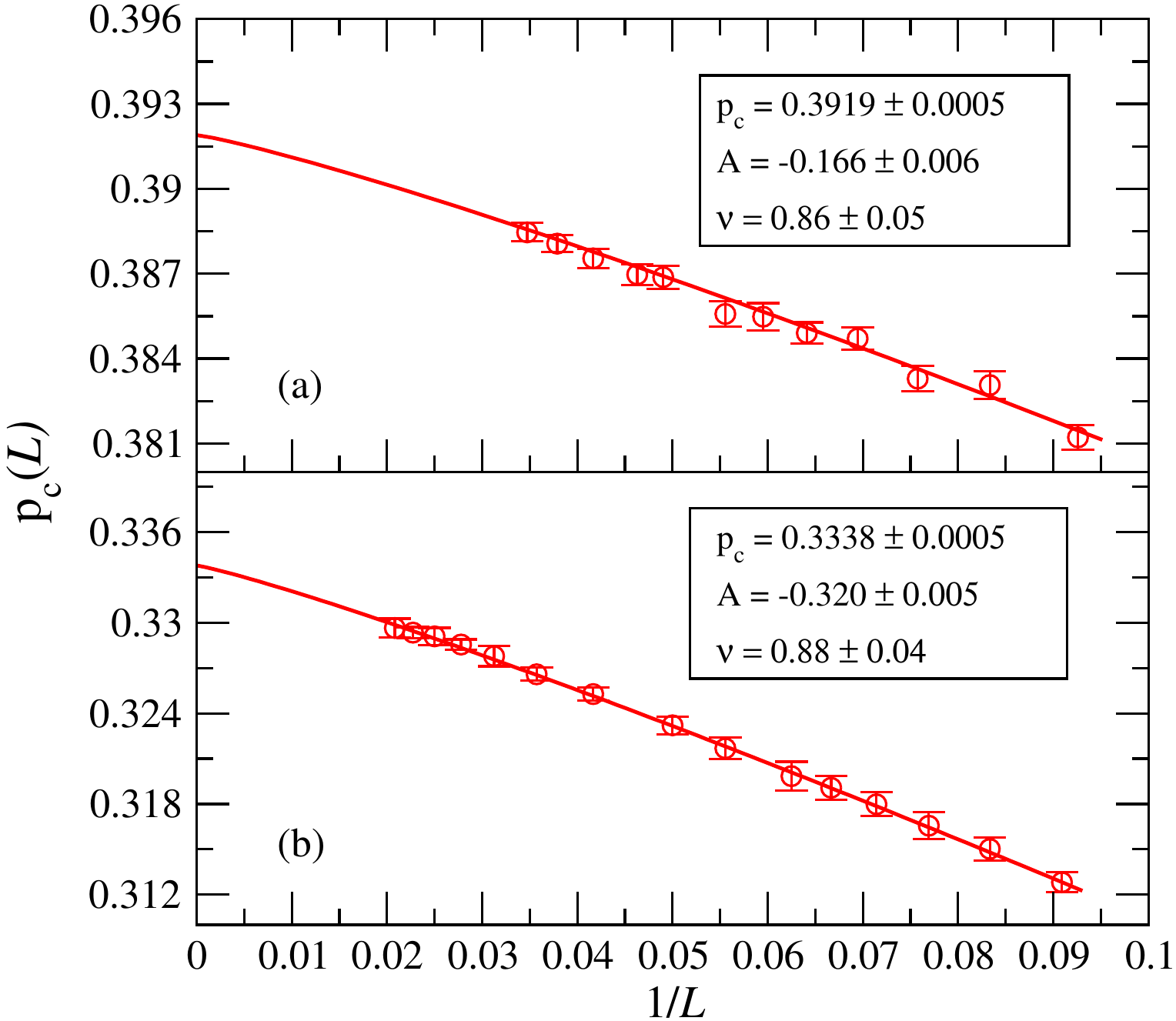}
\centering
\caption{Same as Fig.\,\ref{fig:lieb2d}, but for the three-dimensional layered Lieb lattice. 
}
\label{fig:LLL}
\end{figure}

\begin{figure}
\includegraphics[scale=0.5]{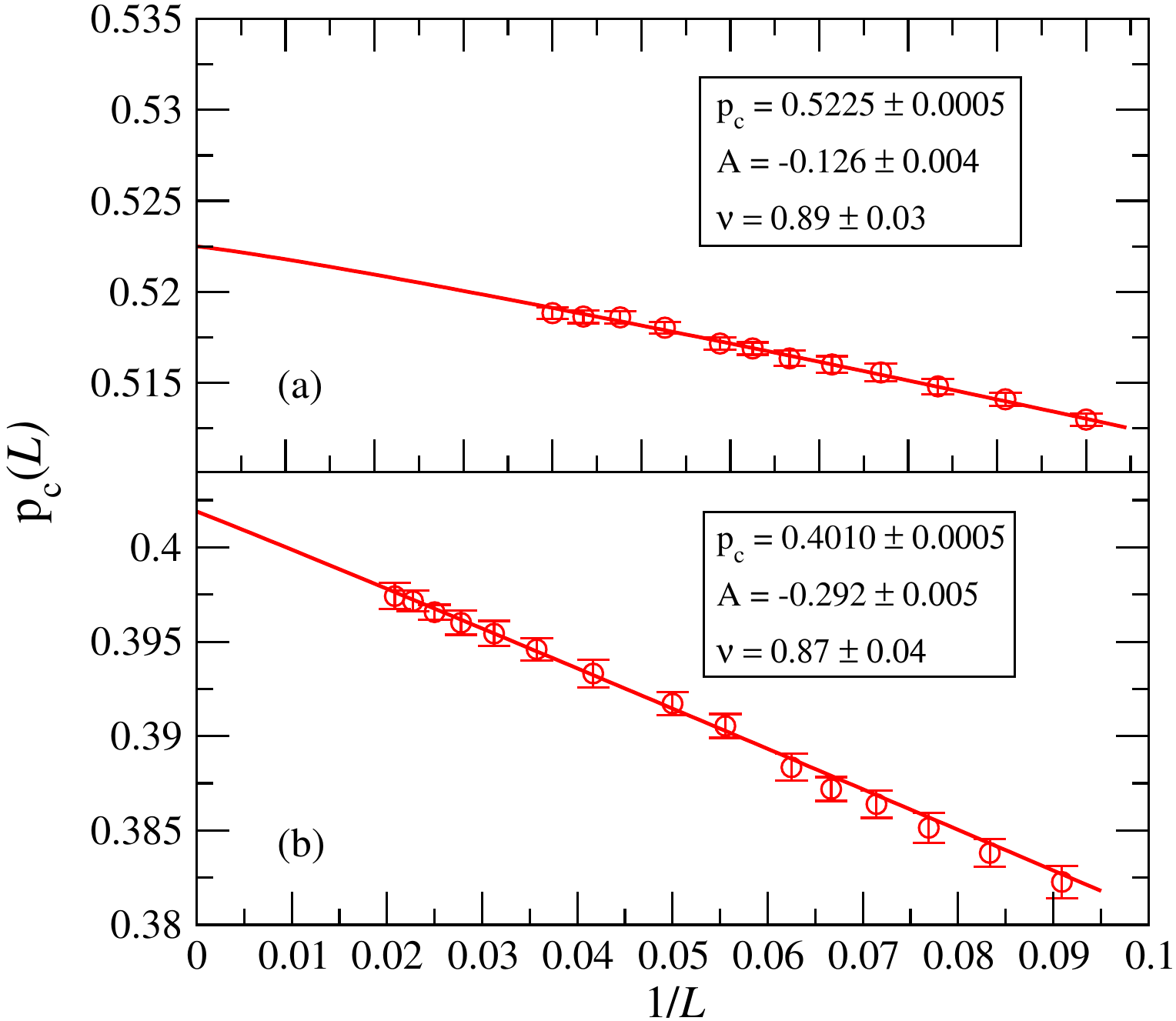}
\centering
\caption{Same as Fig.\,\ref{fig:lieb2d}, but for the three-dimensional perovkite lattice.
}
\label{fig:PL}
\end{figure}

We now discuss the results for the three-dimensional lattices. 
The geometry of the layered Lieb lattice (LLL) we consider consists of $2L$ sites along each of the 
$\hat{\mathbf{x}}$ and $\hat{\mathbf{y}}$ directions, and $L$ layers along the $\hat{\mathbf{z}}$ direction.
We have used between $M=4,000$ runs for the smallest lattices, $L=5$, and $M=800$ for the largest, $L = 28$, lattices. 
The fits to the FSS ansatz, Eq.\,\eqref{eq:FSS}, are shown in Fig.\,\ref{fig:LLL}, the main results of which appear in Table \ref{tab:pcnu}. 
We see that the critical concentrations lie above the corresponding ones for the simple cubic lattice, again due to the smaller connectivity of the O sites in the layers.
The correlation length exponents for the cases of site- and bond-percolation agree, within error bars, with each other and with those for the simple cubic lattice, indicating again that this geometry does not affect the universality class.

Finally, let us discuss the results for the PL, whose geometry is such that one has $2L$ sites along each of the cartesian directions. 
Similarly to the LLL, we have used between $M=4,000$ runs for the smallest lattices, $L=5$, and $M=800$ for the largest, $L = 28$, lattices. 
The fittings to the FSS ansatz, Eq.\,\eqref{eq:FSS}, are shown in Fig.\,\ref{fig:PL}, the main results of which appear in Table \ref{tab:pcnu}. 
We see that the critical concentrations are largest for the PL than for the LLL, since a doubly-coordinated O-site now lies between every six-coordinated Cu-sites, thus leading to the smallest average coordination number of all lattices considered here.
On the other hand, the correlation length exponents for both site- and bond-percolation cases agree, within error bars, with each other and with those for the other 3D lattices.

\section{Conclusions}
\label{sec:conc}

Motivated by recent findings on magnetic and transport properties of fermions in pure and disordered  flat band systems, we have used Monte Carlo simulations to study percolative critical behavior on the two-dimensional Lieb lattice, as well as on its three-dimensional extensions, the layered Lieb lattice and the perovskite lattice.

We have determined thresholds, $p_c$, for both site- and  bond-percolation on these lattices, as well as correlation length exponents.
The accurate values for $p_c$ thus obtained allow us to cast the lattices in order of increasing  $p_c$: 
\begin{equation}
	p_c^\text{sc} < p_c^\text{LLL} < p_c^\text{PL} < p_c^\text{square} < p_c^\text{Lieb},\nonumber  
\end{equation}
which follows the mean-field trend, $p_c^\text{(Bethe)}=(z-1)^{-1}$, with $z$ being interpreted as an average coordination number.
Notwithstanding the fact that each threshold for site-percolation is larger than the one for bond-percolation, the above ordering is applicable to either type of percolation.
The calculated correlation length exponent in each case is consistent with universal behavior for a given lattice dimensionality, $d$; one must therefore expect asymptotic universality in other quantities such as cluster size distribution, and so forth.

\section*{ACKNOWLEDGMENTS}

Financial support from the Brazilian Agencies
CAPES, CNPq and FAPERJ is gratefully acknowledged.

\bibliography{perc_lieb-final.bib}

\end{document}